\def\year{2020}\relax
\documentclass[letterpaper]{article} % DO NOT CHANGE THIS
\usepackage{aaai20}  % DO NOT CHANGE THIS
\usepackage{times}  % DO NOT CHANGE THIS
\usepackage{helvet} % DO NOT CHANGE THIS
\usepackage{courier}  % DO NOT CHANGE THIS
\usepackage[hyphens]{url}  % DO NOT CHANGE THIS
\usepackage{graphicx} % DO NOT CHANGE THIS
\usepackage{subfigure}
\usepackage{amsmath}
\usepackage{graphicx}  % DO NOT CHANGE THIS
\usepackage{bm}
\usepackage[utf8]{inputenc} % allow utf-8 input
\usepackage[T1]{fontenc}    % use 8-bit T1 fonts
\usepackage{hyperref}       % hyperlinks
\usepackage{url}            % simple URL typesetting
\usepackage{booktabs}       % professional-quality tables
\usepackage{amsfonts}       % blackboard math symbols
\usepackage{amsmath}
\usepackage{amssymb}
\usepackage{nicefrac}       % compact symbols for 1/2, etc.
\usepackage{microtype}      % microtypography
\usepackage{adjustbox}
\usepackage{booktabs}
\usepackage{multirow}
\usepackage{adjustbox}
\usepackage{xcolor}
\usepackage{graphics}

\hypersetup{draft}

\newcommand{\ignore}[1]{}

\def\ie{\emph{i.e.}}

\def\etal{\emph{et al.}}

\title{Unsupervised Image Super-Resolution with an Indirect Supervised Path}
\author{Zhen Han$^{2}$, Enyan Dai$^{3}$\thanks{Equal contribution.}, Xu Jia$^{1}$\thanks{Corresponding author}, Xiaoying Ren$^{4}$, Shuaijun Chen$^{1}$, Chunjing Xu$^{1}$,Jianzhuang Liu$^{1}$, Qi Tian$^{1}$ \\
    \normalsize$^1$ Huawei Noah's Ark Lab.\\
    \normalsize$^2$ Institute of statistics and big data, Renmin University of China. \\
    \normalsize$^3$ College of information science and technology, The Pennsylvania State University.\\
    \normalsize$^4$ School of Computer Science and Technology, University of Science and Technology of China.\\
    \small\texttt{handarkholme@ruc.edu.cn;}\\
    \small\texttt{emd5759@psu.edu;}\\
    \small\texttt{\{x.jia,chenshuaijun,xuchunjing,liu.jianzhuang,tian.qi1\}@huawei.com;}\\
    \small\texttt{wjyren@mail.ustc.edu.cn}
}% All authors must be in the same font size and format. Use \Large and \textbf to achieve this result when breaking a line
\begin{document}

\maketitle

\begin{abstract}
  The task of single image super-resolution (SISR) aims at reconstructing a high-resolution (HR) image from a low-resolution (LR) image. Although significant progress has been made by deep learning models, they are trained on synthetic paired data in a supervised way and do not perform well on real data. There are several attempts that directly apply unsupervised image translation models to address such a problem. However, unsupervised low-level vision problem poses more challenge on the accuracy of translation. 
  In this work, we propose a novel framework which is composed of two stages: 1) unsupervised image translation between real LR images and synthetic LR images; 2) supervised super-resolution from approximated real LR images to HR images. It takes the synthetic LR images as a bridge and creates an indirect supervised path from real LR images to HR images. Any existed deep learning based image super-resolution model can be integrated into the second stage of the proposed framework for further improvement. In addition it shows great flexibility in balancing between distortion and perceptual quality under unsupervised setting.   
  The proposed method is evaluated on both NTIRE 2017 and 2018 challenge datasets and achieves favorable performance against supervised methods. 
\end{abstract}

\section{Introduction}
\label{intro}

The task of single image super-resolution (SISR) aims at reconstructing a high-resolution (HR) image from a low-resolution (LR) image. It has been widely used in several computer vision tasks such as image enhancement, surveillance and medical imaging. In the SISR task, an LR image $y$ is modeled as the degradation output by applying the following degradation process to an HR image $x$,
\begin{equation}
 y = \left(x \otimes k\right)\downarrow_s + n,
\end{equation}
where $k$ denotes a blur kernel, $\downarrow_s$ denotes a downsampling operation with scaling factor $s$, and $n$ denotes noise and is usually defined as Gaussian noise with standard deviation $\sigma$.
It is an ill-posed problem since there are multiple solutions that can be reconstructed from a given LR image.

% Xu: This paragraph can be summarized into one or two sentences.
% Most basic methods for SISR are general interpolation based methods, including bilinear interpolation, bicubic interpolation and Lanczos resampling. Such methods do not take image content into consideration and apply a fixed interpolation kernel to an LR image for super-resolving effect without any supervision. 
%
% Recently, learning based, \ie data-driven methods, become popular. These methods can be broadly classified into two families: (i) dictionary based methods~\cite{Chang-cvpr04, Yang-cvpr08, Timofte-iccv13, Timofte-accv14,Yang-eccv14-survey}, and (ii) deep learning based methods~\cite{Dong-tpami16, Shi-cvpr16, Kim-cvpr16-vdsr, Kim-cvpr16-drcn, Tai-cvpr17a, Tai-cvpr17b, Lai-cvpr17, Zhang-cvpr18, Zhang-eccv18}. Both of these two kinds of methods fall into the category of supervised methods.
% In dictionary based methods, a dictionary mapping between LR space and HR space is explicitly learned. Once the mapping is learned, the same set of coding coefficients computed for an LR image are applied to HR space to produce the super-resolved result. 
% Deep learning based methods are built based on the powerful capability of deep neural networks in approximating arbitrary functions. They implicitly learn the mapping between LR and HR images and currently define the state-of-the-art. 

Recently, data-driven methods, especially deep learning based methods~\cite{Dong-tpami16,Shi-cvpr16,Kim-cvpr16-vdsr,Kim-cvpr16-drcn,Tai-cvpr17a,Tai-cvpr17b,Lai-cvpr17,Zhang-cvpr18,Zhang-eccv18} achieve great performance on low-level vision tasks.
Although significant progress have been made on SISR, the trained deep learning models do not perform well on real data. That is because most of them are trained on synthetic paired data in a supervised way. Most synthetic low-resolution images are generated using simple and uniform degradation, such as directly downsampling a high-resolution image with bicubic interpolation. Therefore, the trained models are only able to produce high-quality images on synthetic low-resolution images but with poor generalization ability on unseen, realistic low-resolution images which suffer from other degradation factors such as blur and noise.

There are several works attempting to address this issue. On one hand, some researchers try to improve quality of training data by collecting a dataset consisting of real-world data~\cite{Cai-arxiv19,Zhang-cvpr19}; on the other hand, some researchers work on blind super-resolution~\cite{Glasner-iccv09,Huang-cvpr15,Michaeli-iccv13,Shocher-cvpr18}. These methods work in an unsupervised fashion by exploring self-similarity across scales of an LR image to hallucinate an HR image.
Very recently, ~\cite{Yuan-cvprw18} applies unsupervised image translation to image super-resolution with unpaired LR-HR data. LR images and HR images are considered as two domains and SISR is treated as an LR-to-HR image translation problem. Several recently proposed unsupervised image translation methods such as CycleGAN~\cite{Zhu-iccv17} and DualGAN~\cite{Yi-iccv17} can be employed to super-resolve an image when training data is not paired, showing great potential in addressing super-resolution of unseen, realistic low-resolution images.

Our method shares similar philosophy in the use of cycle consistency to train an SR model for unpaired LR and HR images. 
However, different from the common unsupervised image translation task, there is a strict requirement for the result of image super-resolution. It not only requires the style of a translated image to be correct but also requires that the super-resolved result keeps content and contains few artifacts. That is, it poses more challenge on the accuracy of translation.

The proposed framework is designed particularly for the unsupervised low-level vision problem. 
% It makes full use of characteristics of the SISR task and shows great flexibility in balancing between distortion and perceptual quality.
Instead of directly applying an unsupervised image translation model to bicubic upsampled images or taking a cycle-in-cycle method as in~\cite{Yuan-cvprw18}, we take bicubic interpolated synthetic LR images as an intermediate bridge in the unsupervised training. 
We first build a cycle between real LR space and synthetic LR space. 
By adding a path from real LR images to HR images on top of the path from synthetic LR images to real LR images, we obtain an indirect supervised path from real LR images to HR images. After feeding a synthetic LR image to the cycle model, an LR image with similar degradation to a real LR image can be estimated and an indirect supervised path, synthetic LR images $\rightarrow$ approximated real LR images $\rightarrow$ HR images, can be obtained.
This indirect supervised path allows the proposed method to enjoy advantages of any recently proposed SISR model.
% 1. small size, such that we can use more advanced unsupervised image translation models
% 2. pay more attention to SR models instead of cycle-in-cycle, where it has experience two stages to reach HR from lR.
%
In addition, compared to previous unsupervised image translation approaches, our method is flexible enough such that the proposed real LR-to-HR SR model can be trained with various losses. Unsupervised translation methods trained with cycle-consistency and adversarial loss are prone to produce artifacts because those losses are weak in controlling generation quality. However, for the proposed approach, either a fidelity loss such as L1 or L2, or a perceptual quality oriented loss such as adversarial loss or texture loss can be used for training the proposed models.
Hence, it shows great flexibility in balancing between distortion and perceptual quality.
% Hence, the proposed approach can impose more flexible control on generation quality.
% Finally, we also modify an existed SR network to deal with image-dependent degradation by adding a meta-network as an additional module to modulate the base SR network.
The proposed method is evaluated on both NTIRE 2017 and 2018 challenge datasets and achieves comparable performance to supervised methods. 

% motivation: purely unsupervised training between real LR and HR is prone to produce artifacts because cycle-consistency loss and adversarial loss is weak in cotrolling the generaiton quality; supervised SR model is able to produce much better results but only works for synthetic pairs. whether we can combine adavantages of these two kinds of methods, yes, we can achieve that by using synthetic LR as bridge.

\section{Related Work}
\label{related}

% deep learning based image super-resolution methods
% blind SR ( multiple degradation, Nonparametric blind superresolution, " Zero-Shot" Super-Resolution using Deep Internal Learning, Deep image prior, Patch based blind image super resolution)
% unsupervised image translation methods (cycleGAN, DualGAN, UNIT, StarGAN)
\paragraph{Deep learning based image super-resolution.}
Recently, a lot of works are proposed to address the task of SISR based on Convolutional Neural Networks (CNNs). The first one is proposed by Dong~\etal~\cite{Dong-tpami16}, which implicitly learns a mapping between LR and HR images using a shallow fully-convolutional network. 
% In the Efficient Sub-Pixel Convolutional Neural Network (ESPCN)~\cite{Shi-cvpr16} a sub-pixel layer is added to the end of network, reducing the memory requirement. 
In~\cite{Kim-cvpr16-vdsr,Kim-cvpr16-drcn}, Kim~\etal borrowed the idea of residual connection from ResNet~\cite{He-cvpr16-resnet} and designed a very deep network to improve SISR performance. 
% The upscaling is conducted progressively from a small upscaling factor to a large one in~\cite{Lai-cvpr17}.
Recently, SRResNet~\cite{Caballero-cvpr17}, EDSR~\cite{Lim-cvprws17}, DRRN~\cite{Tai-cvpr17a}, RDN~\cite{Zhang-cvpr18} and ESRGAN~\cite{wang-eccv18ws} proposed to use not only the residual connection in the last layer, but also local residual connections and dense connection in the intermediate layers as in the ResNet~\cite{He-cvpr16-resnet} and DenseNet~\cite{Huang-cvpr17-densenet} architectures, and deeper network architecture to further improve the performance.
In addition, there are several works~\cite{Caballero-cvpr17,Sajjadi-iccv17,wang-eccv18ws} on improving perceptual quality of SISR results by combining fidelity loss with an adversarial loss~\cite{Goodfellow-nips14} and a perceptual loss~\cite{Johnson-eccv16}.
However, they are all trained on synthetic pairs of LR and HR images. LR images in real world suffer from multiple degradations such as complex blur and noise that are difficult to formulate. 
It is difficult and expensive to obtain paired data for real LR and HR images. ~\cite{Cai-arxiv19,Zhang-cvpr19} proposed to capture LR-HR image pairs with a more realistic setting, that is, tuning focal length of DSLR cameras to collect images of different resolution. However, models trained with such data may not generalize well to LR images captured by other devices such as smartphones which may contain different level of noise and blur. 
In this work, we propose an approach to address unpaired data, which allows any aforementioned supervised SR method for synthetic pairs to be trained with unpaired data.

\paragraph{Unsupervised image translation.}
Image super-resolution can be considered as a special image translation task,~\ie, translating images from LR domain to HR domain. 
There have been several approaches to address unsupervised image translation. Zhu~\etal~\cite{Zhu-iccv17} proposed CycleGAN by adding cycle consistency constraint on top of pix2pix~\cite{Isola-cvpr17}. Cycle consistency enforces each image to be correctly reconstructed after translating from a source domain to a target domain and translating back to the source domain. Similar approaches are also proposed in DiscoGAN~\cite{Kim-icml17} and~\cite{Yi-iccv17}. 
Another kind of approaches assume that images from source domain and target domain share a common latent space. Once an image is projected to the shared latent space, a decoder can be used to either reconstruct the image in source domain or produce an image in target domain. Huang~\etal~\cite{Huang-eccv18} and Lee~\etal~\cite{Lee-eccv18} further proposed to decompose an image into a content-related space and a style-related space to achieve many-to-many image translation.
% such that an image can be generated by combining arbitrary content and style representations.

% , hence allowing slight perturbation with translation results. 
% However, super-resolution results must keep the original image's color and content with more details added to enhance its quality. 
%However, super-resolution task has higherq requirement on keeping image's color and content.
%In this work, we add an indirect supervised path to cycled training in order to help impose more control on the translation results.

Inspired by success of these methods, several works propose to use unsupervised image translation methods to model the unpaired mapping between real LR and HR images. \cite{Yuan-cvprw18} introduces two cycles, with one between real LR and synthetic LR images, and another one between real LR and HR images.
\cite{Bulat-eccv18} proposes to first learn a unidirectional GAN-based degradation model for real LR images, and then train an SR model with approximate real LR-HR pairs.
However, these methods attempt to directly translate between real LR and HR images, which may fail due to significant domain gap between real LR and HR images. 
In addition, different from common image translation tasks such as horse-to-zebra translation and style transfer, users are more sensitive to the results of SISR. Models for common image translation do not have strong control on color and content of translation results. 
In this work, we take bicubic interpolated synthetic LR images as an intermediate bridge in the cycled training and construct an indirect supervised path which enables more control on the super-resolution results.
\section{Method}
\label{method}
% Figure for the overall framework refer to CycleGAN's figure 3, and comparison with CycleGAN and supervised method !
% cycleSR - overall idea and architecture
% unsupervised cycle - cycleLR - architecture and its purpose, and advantage
% supervised path - SR model - how to form the supervised path and build bridge between real LR and HR
% loss functions - cycle losses + MSE loss and their relation, texture loss and GAN loss

\begin{figure*}[tp]
	\centering
\begin{tabular}{ccc}		    
        \includegraphics[width=1.0\textwidth]{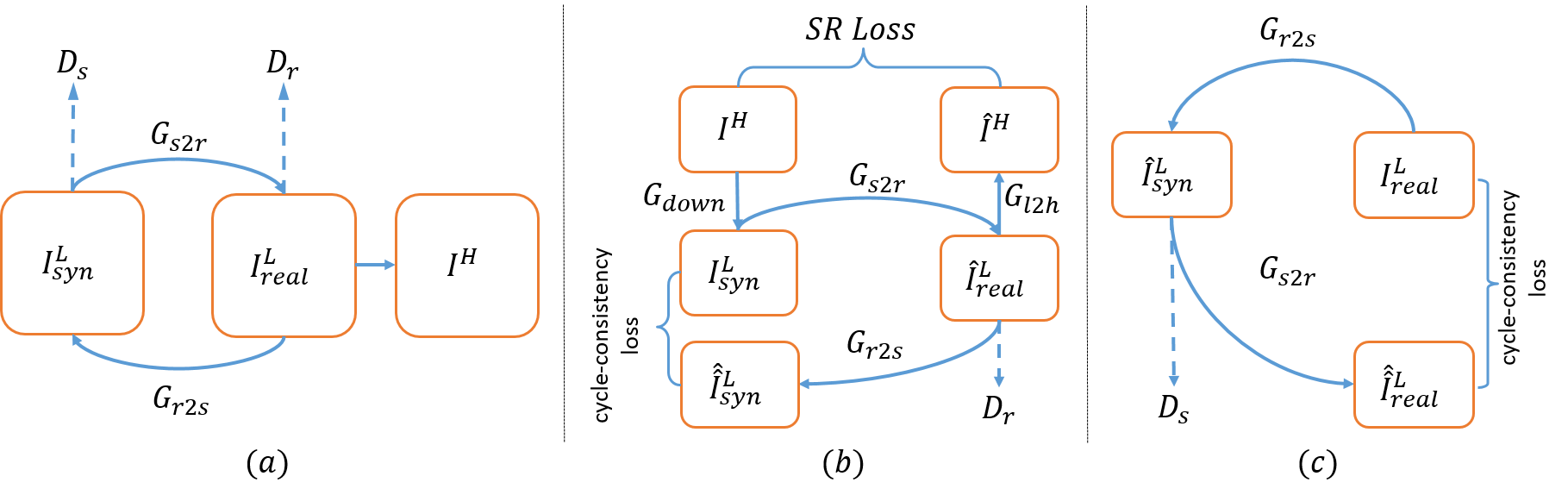}
\end{tabular}
\caption{(a) Pipeline of the proposed framework for unsupervised image super-resolution; (b) forward cycle-consistency and indirect supervised path: $I^L_{syn} \leftrightarrow \hat{I}^L_{real} \rightarrow I^H$; (c) backward cycle-consistency: $I^L_{real} \leftrightarrow \hat{I}^L_{syn}.$}
% \vspace{-3mm}
\label{fig:piepeline}
\end{figure*}

% Different from common unsupervised image translation task, 
As mentioned above, there is strict requirement for the super-resolved result in unsupervised image super-resolution. It not only requires the style of a translated image to be correct but also requires that the super-resolved result keeps the original content and contains few artifacts. Therefore, it poses more challenge on the accuracy of translation.
The proposed framework is designed specifically for unsupervised low-level vision problem and is composed of two stages: 1) unsupervised image translation between real LR images and synthetic LR images; 2) supervised super-resolution from approximated real LR images to HR images. 
Given a real LR image $I^L_{real}$ and an unpaired HR image $I^H$ during training, we first generate a synthetic LR image $I^L_{syn}$ with simple synthetic degradation. Then we have LR images $I^L_{real}$ and $I^L_{syn}$ composing training set for unsupervised translation, and synthetic LR-HR pairs $I^L_{syn}$ and $I^H$ composing training set for supervised image super-resolution. Our goal is to train a model from real LR images $I^L_{real}$ to HR images $I^H$. Assume that a decent unsupervised translation model can succesfully transfer a synthetic LR image $I^L_{syn}$ to an approximated real LR one $\hat{I}^L_{real}$. Then the generated LR images $\hat{I}^L_{real}$ and HR images $I^H$ corresponding to $I^L_{syn}$ implicitly compose supervised training pairs for the task of super-resolving real LR images. An arbitrary SR network proposed for supervised learning can be applied here and be trained with various losses to balance distortion and perceptual quality. The whole framework is shown in Figure~\ref{fig:piepeline}.

\subsection{Unsupervised translation among different degradaded LR images}
% network architecture and loss

In the proposed framework we first take the unsupervised image translation model CycleGAN~\cite{Zhu-iccv17} for mapping between synthetic LR images $I^L_{syn}$ and real LR images $I^L_{real}$. 
Synthetic LR images $I^L_{syn}$ are generated by simply applying bicubic downsampling operation $G_{down}$ to $I^H$ without adding any noise or blur.
In the CycleGAN model, there are a generator $G_{s2r}$ trained to produce samples similar to real LR images and a discrimimnator $D_{r}$ used to distinguish the translated outputs from real LR images. Similarly, there are a generator $G_{r2s}$ trained to learn the mapping for the backward direction and a discriminator $D_{s}$ to detect true synthetic LR images. Cycle consistency, \ie, $G_{r2s}(G_{s2r}(I^L_{syn})) \approx I^L_{syn}$, is added to enable unsupervised learning from synthetic LR images to real LR images. The loss for training such unsupervised image translation model is defined as below. 
\begin{align}
\begin{aligned}
 &L_1(G_{r2s}, D_{s}, G_{s2r}, D_{r}, I^L_{real}, I^L_{syn}) \\
 =& L_{adv}(G_{r2s}, D_{s}, G_{s2r}, D_{r}, I^L_{real}, I^L_{syn}) \\
 +& \lambda_{cyc}L_{cyc}(G_{r2s}, G_{s2r}, I^L_{real}, I^L_{syn})) \\
 +& \lambda_{id}L_{id}(G_{r2s}, G_{s2r}, I^L_{real}, I^L_{syn}),
\end{aligned}
\end{align}
where $L_{adv}(G_{r2s}, D_{s}, G_{s2r}, D_{r}, I^L_{real}, I^L_{syn})$ is the adversarial losses on both real LR domain and synthetic LR domain and contains a term for generators $L^G_{adv}$ and one for discriminators $L^D_{adv}$. In this work, we follow the CycleGAN~\cite{Zhu-iccv17} and use losses in LSGAN~\cite{Mao-iccv17}; $L_{cycle}(G_{r2s}, G_{s2r}, I^L_{real}, I^L_{syn})$ contains cycle consistency losses for both forward and backward directions, and $L_{id}(G_{r2s}, G_{s2r}, I^L_{real}, I^L_{syn})$ includes identity losses for keeping color consistency as proposed in the original CycleGAN. 
%% LSGAN
%\begin{align}
%\begin{aligned}
%    L^G_{adv} =& \mathbb{E}_{x \sim p^{syn}_{\text{data}}}[(D_r(G_{s2r}(x)) - 1)^2] \\
%    +& \mathbb{E}_{y \sim p^{real}_{\text{data}}}[(D_s(G_{r2s}(y)) - 1)^2],
%    \\
%    L^D_{adv} =& \mathbb{E}_{y \sim p^{real}_{\text{data}}}[(D_r(y) - 1)^2] +   \mathbb{E}_{x \sim p^{syn}_{\text{data}}}[(D_r(G_{s2r}(x)))^2] \\
%    +& \mathbb{E}_{x \sim p^{syn}_{\text{data}}}[(D_s(x) - 1)^2] +   \mathbb{E}_{y \sim p^{real}_{\text{data}}}[(D_s(G_{r2s}(y)))^2],
%\end{aligned}
%\end{align}
%%
%% standard GAN
%% \begin{align}
%% \begin{aligned}
%%     L_{adv}(G_{r2s}, D_{s}, G_{s2r}, D_{r}, I^L_{real}, I^L_{syn}) =& \mathbb{E}_{y \sim p^{real}_{\text{data}}}[\log D_r(y)] +   \mathbb{E}_{x \sim p^{syn}_{\text{data}}}[\log (1-D_r(G_{s2r}(x)))] \\
%%     +& \mathbb{E}_{x \sim p^{syn}_{\text{data}}}[\log D_s(x)] +   \mathbb{E}_{y \sim p^{real}_{\text{data}}}[\log (1-D_s(G_{r2s}(y)))],
%% \end{aligned}
%% \end{align}
%%
%\begin{align}
%\begin{aligned}
%    & L_{cyc}(G_{r2s}, G_{s2r}, I^L_{real}, I^L_{syn}) \\
%    =& \ \mathbb{E}_{x \sim p^{syn}_{\text{data}}}[\norm{G_{r2s}(G_{s2r}(x))-x}_1]  \\
%    +& \ \mathbb{E}_{y\sim p^{real}_{\text{data}}}[\norm{G_{s2r}(G_{r2s}(y))-y}_1],
%\end{aligned}
%\end{align}
%%
%\begin{align}
%\begin{aligned}
%    & L_{id}(G_{r2s}, G_{s2r}, I^L_{real}, I^L_{syn}) \\
%    =&  \mathbb{E}_{y \sim p^{real}_{\text{data}}}[\norm{G_{r2s}(y) - y}_1]
%    +& \mathbb{E}_{x \sim p^{syn}_{\text{data}}}[\norm{G_{s2r}(x) - x}_1].
%\end{aligned}
%\end{align}
With a properly trained CycleGAN model, we can obtain an approximate real LR image from a synthetic LR image. The CycleGAN model here does not need to be perfect, because it will be jointly trained with the SR network at the second stage.
Compared to~\cite{Bulat-eccv18} which uses a unidirectional GAN-based High-to-Low network to directly translate HR images to LR ones, our method first obtains bicubic downsampled LR images and then uses a two-way cycle-consistency based network for translation, which obtains more robust translation result than their method.

\subsection{Indirect supervised learning for super-resolution}
% network architecture and loss, including joint losses
Given unpaired real LR images and HR images, there is no pairwise supervision provided to directly train an SR network from real LR images to HR images. However, with the approximated real LR images generated by the CycleGAN from synthetic LR images, we are able to create an indirect supervised path,~\ie, synthetic LR images $\rightarrow$ approximated real LR images $\rightarrow$ HR images. A synthetic LR image is bicubicly downsampled from an HR image and it can be used to produce an approximated real LR image. The approximated real LR image has the same content as the synthetic LR image and HR image, with similar degradation as real LR images. Therefore, we can now indirectly train a supervised SR model $G_{l2h}$ with approximated real LR-HR pairs, which is able to perform well on real LR images.

Note that in this work, we do not explicitly separate the training process of these two stages but combine them together within one framework. 
In this way the SR model is able to adapt to approximated real LR images with slight quality difference. These different results can work as a kind of data augmentation by itself, improving SR model's robustness.
Super-resolution performance of the two-stage model trained separately would be heavily dependent on the result of unsupervised image translation which is still not perfect yet. 
In addition, having these two stages balanced within one framework also further constrains the space of possible mapping function in the CycleGAN. The generated real LR images are required to be not only able to go back to synthetic LR space but also super-resolved to produce the corresponding HR images.
Different from~\cite{Yuan-cvprw18} and~\cite{Bulat-eccv18} where GAN losses plays the major role in recovering noisy LR images, our method shares similar philosophy as supervised SR models and takes distortion measure based losses as main component.
Any loss function that works in the supervised SR training can be used for the indirect supervised path. For example, it can be an MSE loss, a perceptual loss or an adversarial loss. However, it is impossible for a vanilla CycleGAN to be trained with an MSE loss or a perceptual loss since there are no paired output-target data.
The total loss for the proposed framework can be formulated as below.
\begin{align}
\begin{aligned}
  L_{total} &= L_1(G_{r2s}, D_{s}, G_{s2r}, D_{r}, I^L_{real}, I^L_{syn}) \\
  &+ L_2(G_{l2h}, D_h, \hat{I}^L_{real}, I^H),
\end{aligned}
\end{align}
where $L_1(G_{r2s}, D{s}, G_{s2r}, D_{r}, I^L_{real}, I^L_{syn})$ represents the loss of CycleGAN defined above, $L_2(G_{l2h}, D_h, \hat{I}^L_{real}, I^H)$ is SR losses which can be defined as any loss function used to train a supervised SR model. In this work we take a combination of MSE loss, perceptual loss and adversarial loss as our final loss function.
\begin{align}
\begin{aligned}
    L_2(G_{l2h}, D_h, \hat{I}^L_{real}, I^H) = & \lambda_{mse} L_{mse}(G_{l2h},  \hat{I}^L_{real}, I^H) \\
    + & \lambda_{advsr} L_{adv}(G_{l2h}, D_h, \hat{I}^L_{real}, I^H) \\
    + & \lambda_{percep} L_{percep}(G_{l2h}, \hat{I}^L_{real}, I^H),
\end{aligned}
\end{align}
where $\hat{I}^L_{real} = G_{s2r}(I^L_{syn})$ is the approximated real LR images from the first stage, $L_{mse}(G_{l2h}, \hat{I}^L_{real}, I^H)$ and $L_{percep}(G_{l2h}, \hat{I}^L_{real}, I^H)$ are respectively reconstruction loss and perceptual loss between super-resolved results and HR images, and $L_{adv}(G_{l2h}, D_h, \hat{I}^L_{real}, I^H)$ is adversarial loss on HR space, with a term $L^G_{advsr}$ for the generator and a term $L^D_{advsr}$ for the discriminator. We follow the ESRGAN~\cite{wang-eccvws18} and use losses in Relativistic GAN~\cite{RaGAN-iclr19}. 

Moreover, another advantage of the proposed framework over the vanilla CycleGAN is that the cycle training in our framework is conducted on LR space. During the cycled training, there are two generators and two discriminators to be trained. Therefore, plenty of GPU memory is required if the training is conducted on HR space. Besides, the advantage over other two-stage methods like CinCGAN~\cite{Yuan-cvprw18} lies in that pipeline of those methods is also two-stage during inference. However, the proposed approach is only two-stage during training but one-stage during inference. Therefore, the proposed approach takes up much less memory and runs faster during inference than other SR methods with unsupervised translation model.

% \subsection{Meta-network}

% treat it as an autoencoder?

\section{Experiments}
\label{experiment}
% dataset (NTIRE 2017 unknown degradataion dataset, NTIRE 2018 mild and wild degradataion dataset, number of images, and degradation)
% training details including data preparation and network training
% ablation study 
% (1) bicubic interpolation, synthetic paired data, paired data, cycleGAN+SR, the proposed method
% (2) influence of loss weights
% (3) perceptual quality
% (the proposed method with different weights on NTIRE 2017 unknown degradataion dataset and NTIRE 2018 mild degradation dataset)
In this section, we first introduce the datasets for evaluation under unpaired image super-resolution setting. Note that there is no unpaired dataset with real LR and HR images. In this work, we treat LR images with advanced degradation as real LR images and will explain how to form unpaired training data later. Implementation details for training the proposed models are described afterwards. Finally, we compare the proposed approach with several baselines, and investigate the influence of loss weights at the second stage on the super-resolution performance. 

\subsection{Data}
DIV2K~\cite{NTIRE17,NTIRE18} is a popular SISR dataset which contains 1,000 images with different scenes and is splitted to 800 for training, 100 for validation and 100 for testing. It was collected for NTIRE2017 and NTIRE2018 Super-Resolution Challenges in order to encourage research on image super-resolution with more realistic degradation. This dataset contains LR images with different types of degradations.
% We applied our method on both NTIRE 2017 \cite{NTIRE17} and 2018 challenge datasets \cite{NTIRE18}. Both challenges use the DIV2K dataset \cite{NTIRE17}, which contains 1000 high resolution images with various contents. 
Apart from the standard bicubic downsampling, several types of degradations are considered in synthesizing LR images for different tracks of the challenges. Since this work aims at dealing with more realistic setting, we choose track 2 of NTIRE 2017 which contains LR images with unknown x4 downscaling, and track 2 and track 4 of NTIRE 2018 which respectively correspond to realistic mild $\times4$ and realistic wild $\times4$ adverse conditions. 
More specifically, LR images under realistic mild x4 setting suffer from motion blur,
Poisson noise and pixel shifting; while degradations under realistic wild x4 setting are further extended to be of different levels from image to image, which is more realistic and challenging.
DIV2K dataset is split to training set, validation set and test set. Our models are trained on the 800 training images. In our experiment, since we focus on dealing with LR images with unpaired HR images, we take the first 400 HR images in the traning set as training data for HR image domain, and LR images corresponding to the other half for LR image domain. It is evaluated on the whole validation set of 100 images. Due to the motion blur with degraded images, input LR images and ground truth HR images are not well aligned. Following~\cite{NTIRE18}, we ignore border pixels and report the most favorable score among images shifted from 0 to 40 pixels. For the two tracks in NTIRE 2018, a $60\times60$ centered image patch is cropped to compute scores.

\begin{table*}[!tbp]
	\centering	
	\caption{Quantitative comparison for $4\times$ SR on three datasets: average PSNR/SSIM for scale factor x4. 
		{\color{blue}{\textbf{Blue}}} text indicates the best and {\color{green}{\textbf{green}}} text indicates the second best performance.}
	\scalebox{1.0} 
	{
		\begin{tabular}{lccccccc}
			\toprule
			Datasets & Bicubic & SR\_syn & SR\_paired & CycleGAN & Cycle+SR & CycleSR & CycleSRGAN % & Ours\_mse & Ours\_gan
			\\
			\midrule    
			NTIRE17 T2 & 23.976/0.644 & 23.955/0.654 & {\color{blue}{\textbf{29.819}}}/{\color{blue}{\textbf{0.818}}} & 23.213/0.648 & 24.663/0.685 & {\color{green}{\textbf{27.021}}}/{\color{green}{\textbf{0.770}}} & 25.978/0.737
			\\
			NTIRE18 T2 & 23.196/0.563 & 23.066/0.545 & {\color{green}{\textbf{24.133}}}/{\color{green}{\textbf{0.616}}} & 22.901/0.517 & 23.603/0.612 & {\color{blue}{\textbf{24.779}}}/{\color{blue}{\textbf{0.631}}} & 23.605/0.545
			\\
			NTIRE18 T4 & 22.579/0.543 & 22.410/0.517 & {\color{green}{\textbf{23.697}}}/{\color{green}{\textbf{0.590}}} & 21.685/0.466 & 23.357/0.582 & {\color{blue}{\textbf{23.807}}}/{\color{blue}{\textbf{0.593}}} & 22.303/0.497
			\\
			\bottomrule
		\end{tabular}
	}
	%  \vspace{-5mm}
	\label{tab:quantitative}
\end{table*}

\subsection{Implementation details}

\textbf{Network Architecture.} 
Our framework is composed of two stages,~\ie, unsupervised image translation and supervised super-resolution. As for the first stage, we take the network of CycleGAN \cite{Zhu-iccv17}for translation between real LR images and synthetic LR images; as for the super-resolution network, we take different network architectures for different tracks. As for track 2 of NTIRE 2017, we adopt a modified version of VDSR~\cite{Kim-cvpr16-vdsr} by removing the bicubic upsampling operation at the beginning, adding a pixel shuffling layer~\cite{Shi-cvpr16} at the end, and replacing each of 20 convolutional layers with a BN-ReLU-Conv block. As for the track 2 and 4 of NTIRE 2018, we adopt the SRResNet~\cite{Caballero-cvpr17} instead. The reason lies in that the two tracks of NTIRE 2018 suffer from severe motion blur up to 40 pixel shift and more severe noise. Therefore, a deeper network with larger receptive field is required.
% For the unknown dataset, we use 20 convolution layers with batch normalization as the major body of the network. For the mild and wild dataset, we use 16 residual blocks. Because the motion blur is much stronger, which could cause the shift up to 40 pixels, it is necessary to use a deeper network with larger receptive field. Relativistic discriminator \cite{wang2018esrgan} is adopted when we train our model for better perceptual quality.  
%
\begin{figure*}[!t]
	\scriptsize
	\centering
	\begin{tabular}{cc}
		% scriptsize and scalebox violates paper checker
		\begin{adjustbox}{valign=t}
			\begin{tabular}{c}
				\includegraphics[height=0.16\textwidth]{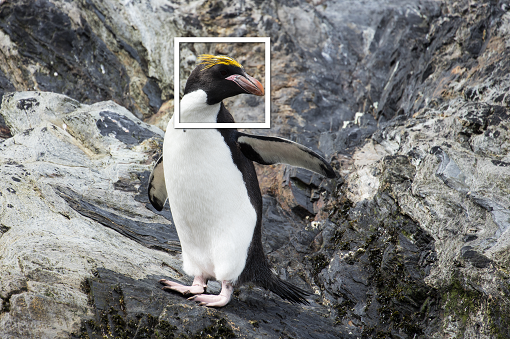}
				\\
				0801 HR 
				\\
				from NTIRE 2017 Track 2
			\end{tabular}
		\end{adjustbox}
		\begin{adjustbox}{valign=t}
			\begin{tabular}{c@{\hspace{1mm}}c@{\hspace{1mm}}c@{\hspace{1mm}}c}
				\includegraphics[width=0.15\textwidth]{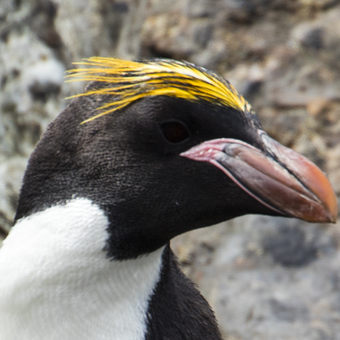}
				&
				\includegraphics[width=0.15\textwidth]{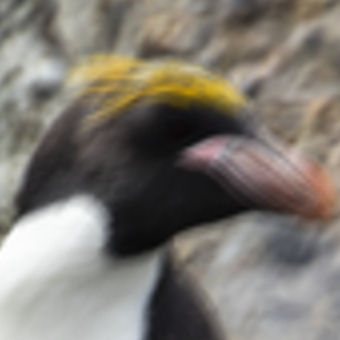}
				&
				\includegraphics[width=0.15\textwidth]{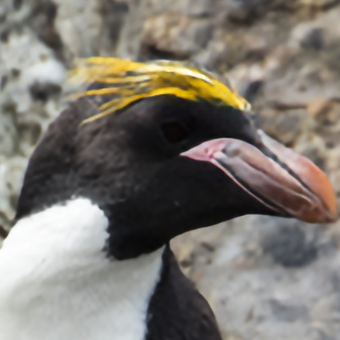}
				&
				\includegraphics[width=0.15\textwidth]{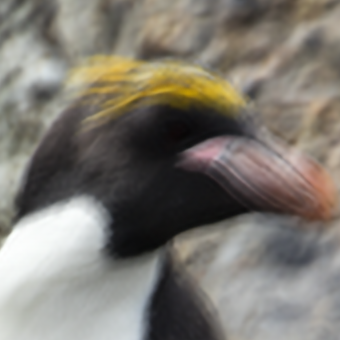}
				\\
				% 				HR (PSNR, SSIM) &
				% 				Bicubic (23.57, 0.69) &
				% 				SR Paired (30.92, 0.89) &
				% 				Synthetic SR (23.53, 0.69)		
				%
				HR &
				Bicubic  &
				\textbf{SR Paired} &
				Synthetic SR 
				\\
				(PSNR, SSIM)
				&
				(23.57, 0.69)
				&
				(\textbf{30.92}, \textbf{0.89})
				&
				(23.53, 0.69)
				\\
				\includegraphics[width=0.15\textwidth]{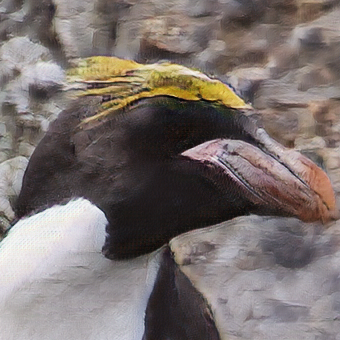}
				&
				\includegraphics[width=0.15\textwidth]{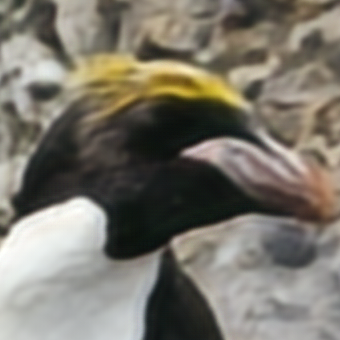} 
				&
				\includegraphics[width=0.15\textwidth]{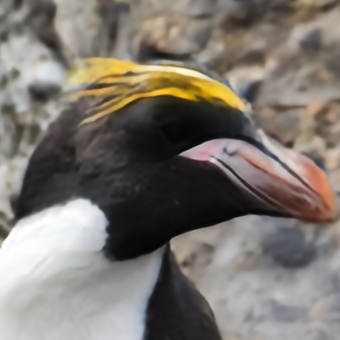} 
				&
				\includegraphics[width=0.15\textwidth]{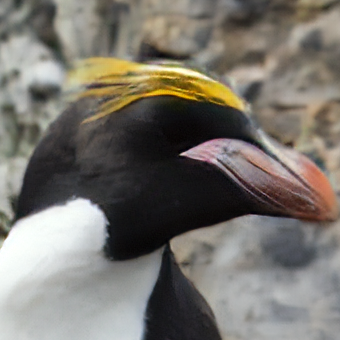} 
				\\ 
				% 				CycleGAN (23.64, 0.66) &
				% 				Cycle+SR (25.23, 0.75) &
				% 				CycleSR  (26.74, 0.83) &
				% 				CycleSRGAN (26.25, 0.81) 
				%
				CycleGAN  &
				Cycle+SR  &
				\textbf{CycleSR}  &
				CycleSRGAN 
				\\
				(23.64, 0.66)
				&
				(25.23, 0.75)
				&
				(\textbf{26.74}, \textbf{0.83})
				&
				(26.25, 0.81) 
			\end{tabular}
		\end{adjustbox}
	\end{tabular}

	\begin{tabular}{cc}
		\begin{adjustbox}{valign=t}
			\begin{tabular}{c}
				\includegraphics[height=0.16\textwidth]{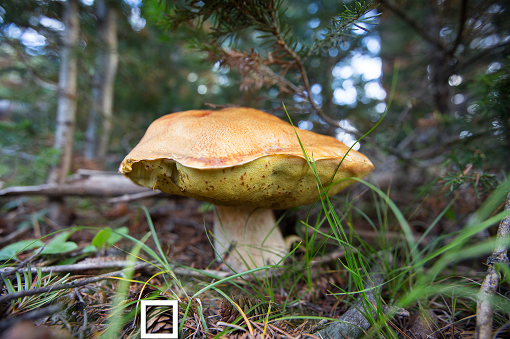}
				\\
				0815 HR 
				\\
				from NTIRE 2018 Track 2
			\end{tabular}
		\end{adjustbox}
		\begin{adjustbox}{valign=t}
			\begin{tabular}{c@{\hspace{1mm}}c@{\hspace{1mm}}c@{\hspace{1mm}}c}
				\includegraphics[width=0.15\textwidth]{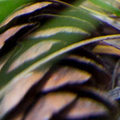}
				&
				\includegraphics[width=0.15\textwidth]{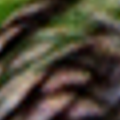}
				&
				\includegraphics[width=0.15\textwidth]{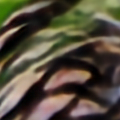}
				&
				\includegraphics[width=0.15\textwidth]{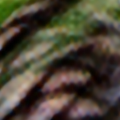}
				\\
				HR &
				Bicubic &
				\textbf{SR Paired} &
				Synthetic SR 
				\\
				(PSNR, SSIM)
				&
				(20.85, 0.61) 
				&
				(\textbf{23.77}, \textbf{0.72})
				&
				(21.12, 0.63)
				\\
				\includegraphics[width=0.15\textwidth]{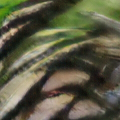}
				&
				\includegraphics[width=0.15\textwidth]{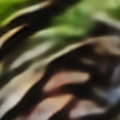} 
				&
				\includegraphics[width=0.15\textwidth]{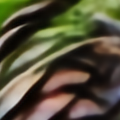} 
				&
				\includegraphics[width=0.15\textwidth]{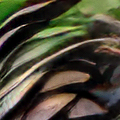} 
				\\ 
				CycleGAN &
				Cycle+SR &
				\textbf{CycleSR} &
				CycleSRGAN 
				\\
				(22.00, 0.65) 
				&
				(23.47, 0.71) 
				&
				(\textbf{23.94}, \textbf{0.74}) 
				&
				(20.90, 0.62)
			\end{tabular}
		\end{adjustbox}
	\end{tabular}
	\begin{tabular}{cc}
		\begin{adjustbox}{valign=t}
			\begin{tabular}{c}
				\includegraphics[height=0.16\textwidth]{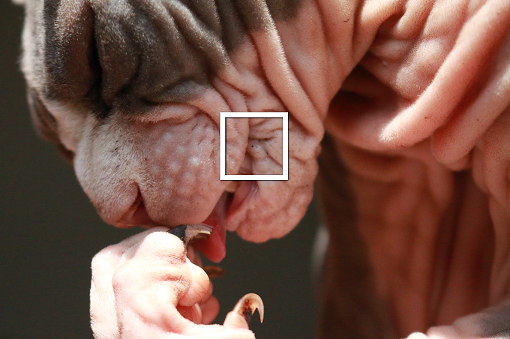}
				\\
				0838 HR 
				\\
				from NTIRE 2018 Track 4
			\end{tabular}
		\end{adjustbox}
		\begin{adjustbox}{valign=t}
			\begin{tabular}{c@{\hspace{1mm}}c@{\hspace{1mm}}c@{\hspace{1mm}}c}
				\includegraphics[width=0.15\textwidth]{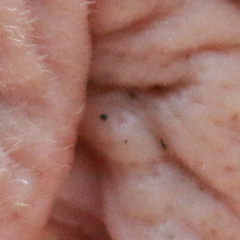}
				&
				\includegraphics[width=0.15\textwidth]{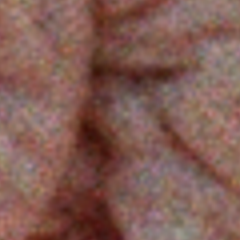}
				&
				\includegraphics[width=0.15\textwidth]{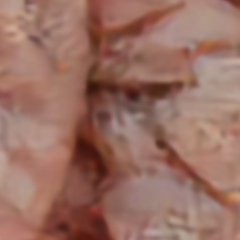}
				&
				\includegraphics[width=0.15\textwidth]{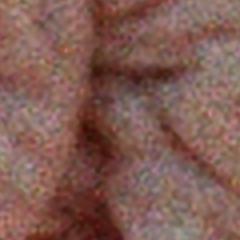}
				\\
				HR &
				Bicubic &
				\textbf{SR Paired} &
				Synthetic SR 
				\\
				(PSNR, SSIM)
				&
				(21.20, 0.78)
				&
				(\textbf{29.54}, \textbf{0.82})
				&
				(20.92, 0.71)
				\\
				\includegraphics[width=0.15\textwidth]{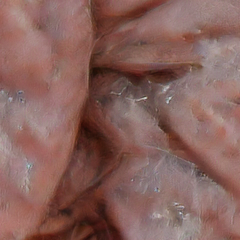}
				&
				\includegraphics[width=0.15\textwidth]{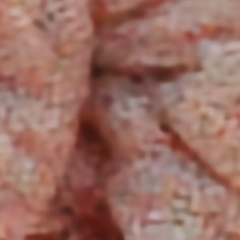} 
				&
				\includegraphics[width=0.15\textwidth]{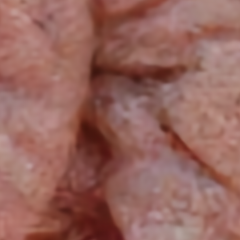} 
				&
				\includegraphics[width=0.15\textwidth]{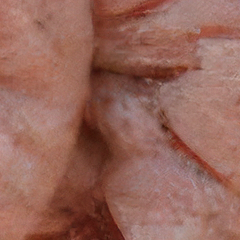} 
				\\ 
				CycleGAN &
				Cycle+SR &
				\textbf{CycleSR} &
				CycleSRGAN 
				\\
				(23.29, 0.74) 
				&
				(26.45, 0.81) 
				&
				(\textbf{28.37}, \textbf{0.86}) 
				&
				(26.14, 0.74) 
			\end{tabular}
		\end{adjustbox}
	\end{tabular}
	\vspace{-0.2cm}
	\caption{Visual comparison for $4\times$ SR on three datasets. The best two results are in \textbf{bold}}
	% 	\vspace{-2mm}
	\label{fig:qualitative}
\end{figure*}
\\
\\
\textbf{Training details.} 
We crop $120\times120$ patches from each LR/HR training image and obtain around 33000 patchs in total. They are flipped and rotated on the fly for further data augmentation. In this work bicubic down-sampling is used to generate synthetic LR images and it can be replaced by other downsampling methods such as nearest neighbor and bilinear down-sampling. The mini-batch size is set to 32. The weights for loss terms in the CycleGAN are set to $\lambda_{cyc} = 10, \lambda_{id} = 0.5$ for all datasets. As for the weights for SR loss terms, $\lambda_{percep}$ and $\lambda_{advsr} = 0.05$ are respectively set to 1 and 0.05 for all datasets, while $\lambda_{mse}$ is dependent on degradation of the dataset, with 1e3 for the track2 of NTIRE2018 and 1e2 for the tracks of NTIRE2018. We use Adam \cite{Kingma-iclr15-adam} optimizer with $\beta_1 = 0.9$, $\beta_2 = 0.999$. The learning rate is set to $2\times10^{-4}$ and $1\times10^{-4}$ for the CycleGAN and SR network. After traning for 100 epoches, the learning rate starts to linearly decay and stops at zero after another 100 epochs. To stablize the adversarial training, we restrict the norm of gradient up to 50. In addition, during training we first pretrain a CycleGAN on unpaired LR images and an SR network on synthetic LR-HR pairs respectively for 5 epochs and then jointly train the whole framework till 200 epochs. Pretraining provides a good initialization for the whole framework and helps stablize the training.

% Start

\subsection{Results}
In this section, we compare the proposed method with several baseline methods on three datasets. Both quantitative and qualitative results are shown to demonstrate its effectiveness. We denote the proposed method trained with only MSE as \textit{CycleSR} and the one trained with both MSE and perceptual quality oriented losses as \textit{CycleSRGAN}.
\paragraph{Baselines.} 
\textit{Bicubic}: bicubic interpolation works without supervision and can be applied to any case;
\textit{SR\_syn}: we train the same SR network on synthetic LR-HR image pairs and evaluate it on the datasets correspond to the three tracks;
\textit{SR\_paired}: we also train the same SR network on real LR-HR image pairs, with a preprocessing of alignment for shifted pixels caused by severe blur in case of Track 2 and Track 4 in NTIRE 2018.
%\textit{SR\_paired}: we also train the same SR network directly on the paired LR-HR images without any preprocessing as a kind of oracle;
\textit{CycleGAN}: we first resize an LR image to the target size and directly apply CycleGAN to the unpaired data;
\textit{Cycle+SR}: we take the same network architecture as the proposed method but train two stages separately.

\paragraph{Quantitative results.}
We present PSNR and SSIM scores for each compared method in Table~\ref{tab:quantitative}.
As shown in Table~\ref{tab:quantitative}, the method trained on synthetic pairs performs even worse than the simple bicubic upsampling on all three datasets, which implies the importance of study on image super-resolution with unpaired data. 
%The supervised method still outperforms unsupervised methods on NTIRE 2017 Track 2 but fails to produce good scores on tracks of NTIRE 2018. 
Among those methods, the best results in all three datasets are taken by either the supervised method or ours. The supervised method outperforms ours on Track 2 of NTIRE 2017, while ours are better on both tracks of NTIRE 2018.
%That is because images in NTIRE 2017 Track2 suffer less from motion blur but the ones in the two tracks of NTIRE 2018 contain severe blur. 
That is, in case of severe blur and much noise, the proposed method CycleSR performs even better than the supervised method which is trained on pre-aligned input-target pairs. The supervised methods which rank high on the leaderboard of NTIRE 2018 also take image registration as a preprocessing step and use deeper network architecture. 
The best three methods for track 2 are \textit{nmhkahn} and \textit{iim\_lab} with $24.73/0.72$, $24.59/0.60$ and $24.59/0.60$, and those for track 4 are \textit{xixihaha} and \textit{yyuan13} with $24.12/0.56$, $24.07/0.56$ and $23.90/0.56$. 
% Although it is not possible for unsupervised methods to conduct such preprocessing, 
Without any such preprocessing, the proposed method CycleSR still achieves favorable performance against those top ranked methods.
The method with the same network architecture but with the two stages trained separately also obtains reasonable performance, but performs worse than the one trained jointly.
In addition, due to the flexibility of the proposed framework, it can be easily trained with perceptual quality oriented losses. Not only can the proposed method obtain high scores but can also produce images of good perceptual quality, which would be discussed later. 
Overall, the proposed unsupervised method are better than other unsupervised baseline methods and is competitive even compared to supervised ones.

\begin{figure*}[!t]
	\scriptsize
	\centering
	\scalebox{0.9} {
		% \begin{tabular}{c@{\hspace{0.001\linewidth}}c@{\hspace{0.001\linewidth}}c@{\hspace{0.001\linewidth}}c}			    
		\begin{tabular}{c@{\hspace{1mm}}c@{\hspace{1mm}}c@{\hspace{1mm}}c@{\hspace{1mm}}c@{\hspace{1mm}}c@{\hspace{1mm}}c}			    
			\includegraphics[width=0.15\textwidth]{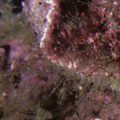}
			&
			\includegraphics[width=0.15\textwidth]{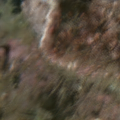}
			&
			\includegraphics[width=0.15\textwidth]{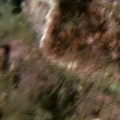}
			&
			\includegraphics[width=0.15\textwidth]{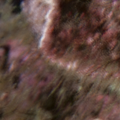}
			&
			\includegraphics[width=0.15\textwidth]{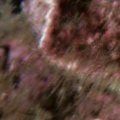}
			&
			\includegraphics[width=0.15\textwidth]{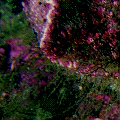}
			&
			\includegraphics[width=0.15\textwidth]{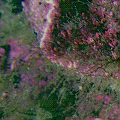}
			\\
			$I^L_{syn}$
			&
			$\hat{I}^L_{real}$
			&
			$\hat{I}^L_{real norm}$
			&
			$\hat{I}^L_{real}$
			&
			$\hat{I}^L_{real norm}$
			&
			$\hat{I}^L_{real}$
			&
			$\hat{I}^L_{real norm}$ 
			\\
			~ & \multicolumn{2}{c}{$\lambda_{mse} = 100$} & \multicolumn{2}{c}{$\lambda_{mse} = 1000$} & \multicolumn{2}{c}{$\lambda_{mse} = 10000$}
		\end{tabular} 
	}
	\caption{Approximated LR images generated with different $\lambda_{mse}$ on NTIRE17 Track 2.}
	\label{fig:l_real_hat}	
\end{figure*}

\paragraph{Qualitative results.}
As shown in Figure~\ref{fig:qualitative}, %the proposed method and supervised method both show better perceptual quality than others. 
the SR network trained on synthetic pairs only learns to sharpen textures but cannot deal with motion blur or noise, while the same network trained on pre-aligned real LR-HR pairs achieves much better visual results in terms of sharpening and denoising. 
%The results of SR network directly trained on paired data on NTIRE 2018 look very blurry, which implies it is difficult for an SR network to deal with both severe misalignment and noise simultaneously. 
The proposed CycleSR also obtains comparable performance as SR Paired. However, SR Paired and such supervised method rely heavily on aligned real LR-HR pairs. In other words, the result of SR Paired is determined by the pre-aligned step of data, while our CycleSR does not need any preprocessing and achieves favourable results as well. 
%The proposed method does not learn to explicitly address these two issue. Instead, it learns to remove blur and noise with adversarial learning and cycle consistency constraint. 
Compared to CycleGAN, the proposed CycleSRGAN is able to produce high quality images with less artifacts and distortion. For all three examples in Figure~\ref{fig:qualitative}, CycleGAN changes the color of translated results and adds many unexisted texture patterns. From the second example in the figure, we can see that CycleSRGAN produces an image even sharper than the ground truth HR image. That is because that HR image is taken with focus on the centered mushroom. The pine cone in the image belongs to out-of-focus region. Our method does not take depth and focus into consideration and tries to recover all blurred parts.

\paragraph{Influence of indirect supervised path:}
We also investigate the influence of loss weights at the second stage on the training of the proposed method, especially for the CycleSR model. 
The weight of MSE loss at the second stage $\lambda_{mse}$ is chosen from $\{1e2, 1e3, 1e4\}$.
Figure~\ref{fig:l_real_hat} shows approximated LR images $\hat{I}^L_{real}$,~\ie, intermediate output of the first stage, trained with different weights of MSE loss at the second stage. Since range of pixel values in $\hat{I}^L_{real}$ varies with different $\lambda_{mse}$, we normalize $\hat{I}^L_{real}$ in terms of the mean and variance of $\hat{I}^L_{syn}$ to get $\hat{I}^L_{real norm}$ for better visualization.
When $\lambda_{mse}$ is small with $\lambda_{mse}=1e2$, CycleGAN weighs more
% plays a more important role 
in the training of the whole framework. In this case, $\hat{I}^L_{real}$ has similar style as real images $I^L_{real}$, but suffers from color drift like CycleGAN. The model trained only learns to reconstruct the correct HR image from color-drifted LR image and would fail during inference phase.
As $\lambda_{mse}$ increases to $\lambda_{mse}=1e3$, color drift problem becomes less severe. The MSE loss at the second stage works as a regularizer to help translation.
When it is too big with $\lambda_{mse}=1e4$, the MSE loss at the second stage dominates the whole optimization process. Instead of generating an image similar to a real LR image, the generator at the first stage tries to generate an image which is easier for the SR network at the second stage to obtain good reconstruction. The intermediate output $\hat{I}^L_{real}$ looks like synthetic LR images rather than real LR images. Hence it would not work to reconstruct a HR image given a real LR image. We also evaluate these variants on NTIRE2017 Track2. The results for $\lambda_{mse} = 1e2, 1e3, 1e4$ are respectively 26.262/0.735, 27.021/0.77, 18.337/0.599, which is consistent with our analysis before. 
% Therefore, 
It is important to balance between the losses of the two stages of the proposed method for good performance.

\section{Conclusion}
\label{conclusion}
% Although significant progress has been made by deep learning methods, most of them only work on synthetic paired data and are limited in generalizing to real data. 
In this work, we present a general framework for unsupervised image super-resolution, which is closer to real scenario. Instead of directly applying unsupervised image translation to address this task, we propose a novel approach which integrates cycled training and supervised training into one framework. 
% The proposed approch consists of two stages: unsupervised image translation between real LR images and synthetic LR images; supervised super-resolution from approximated real LR images to HR images. 
Synthetic LR images are taken as a bridge and creates an indirect supervised path from real LR images to HR images. 
We show that the proposed approach learns to super-resolve a real LR image without any corresponding HR images in the training dataset. It is flexible enough to integrate any existed deep learning based super-resolution models, including those trained with either fidelity losses or perceptual oriented losses. It is evaluated on image super-resolution challenge datasets and achieves favorable performance against supervised methods.

%\clearpage

\small
\bibliographystyle{./aaai}
\bibliography{./ref}

\end{document}